\documentclass[prc,twocolumn,showpacs,preprintnumbers,amsmath,amssymb,superscriptaddress]{revtex4}

\usepackage{graphicx} 
\usepackage{dcolumn}  
\usepackage{bm}       

\def\be{\begin{equation}}
\def\ee{\end{equation}}

\def\delo{$\Delta^{(3)}_o$}

\newcommand{\ev}{\sc ev8}
\newcommand{\hfbtho}{\sc hfbtho}
\begin{document}

\title{Odd-even mass differences from self-consistent mean field theory}

\author{G.F. Bertsch}
\affiliation{Institute for Nuclear Theory and Dept. of Physics,
University of Washington, Seattle, Washington}
\author{C.A. Bertulani}
\affiliation{Department of Physics, Texas A\&M University-Commerce, Commerce, Texas 75429, USA}
\author{W. Nazarewicz}
\affiliation{Department of Physics and Astronomy, University of
Tennessee, Knoxville, TN 37996, USA}
\affiliation{Oak Ridge National
Laboratory, P.O. Box 2008, Oak Ridge, TN 37831, USA}
\affiliation{Institute of Theoretical Physics, Warsaw University,
ul.Ho\.{z}a 69, PL-00681 Warsaw, Poland}
\author{N. Schunck}
\affiliation{Department of Physics and Astronomy, University of
Tennessee, Knoxville, TN 37996, USA}
\affiliation{Oak Ridge National
Laboratory, P.O. Box 2008, Oak Ridge, TN 37831, USA}
\author{M.V. Stoitsov}
\affiliation{Department of Physics and Astronomy, University of
Tennessee, Knoxville, TN 37996, USA}
\affiliation{Oak Ridge National
Laboratory, P.O. Box 2008, Oak Ridge, TN 37831, USA}
\affiliation{Institute of Nuclear Research and Nuclear Energy,
Bulgarian Academy of Sciences, Sofia, Bulgaria}

\date{INT PUB08-52}

\begin{abstract}

We survey odd-even nuclear binding energy staggering using 
density 
functional theory with several treatments of the pairing interaction
including the BCS, Hartree-Fock-Bogoliubov,
and the Hartree-Fock-Bogoliubov with the Lipkin-Nogami approximation.
We calculate the second difference of binding energies and compare with
443 measured neutron energy differences
in isotope chains and 418 measured proton energy differences in isotone 
chains.  The particle-hole part of the energy functional is taken as the 
SLy4 Skyrme parametrization and the pairing part of the functional 
is based on a contact interaction with possible density dependence.
An important feature of the data, reproduced by the theory, is the
sharp gap quenching at magic numbers. 
With the strength of the interaction as a free parameter, the theory
can reproduce the data to an rms accuracy of about 0.25 MeV.  This is
slightly better than a single-parameter phenomenological description but
slightly poorer than the usual two-parameter phenomenological form
$C/A^\alpha$.  
The following conclusions can be made about the 
performance of common parametrization of the pairing interaction:
(i) there is a weak preference
for a surface-peaked neutron-neutron pairing, which might be attributable
to many-body effects; (ii) a larger strength is required
  in the proton pairing
channel than in the neutron pairing channel; (iii)  pairing strengths
adjusted to the well-known 
spherical
isotope chains are too weak to give a good
overall fit to the mass differences.
\end{abstract}


\maketitle

\section{Introduction}

The theory of nuclear masses or binding energies has attracted renewed
interest with the advent of computational resources sufficient to 
performed global calculations based on self-consistent mean-field theory,
also called density functional theory (DFT)  \cite{[Ben03w],[Sto06],[Ber07]}.  
A long-term goal is an 
improved reliability for a theory that avoids {\it ad hoc} phenomenological
parametrizations.  One particular aspect of the nuclear binding
problem is the ubiquitous phenomenon of odd-even  staggering
(OES) of binding energy.
Since the early days of BCS theory \cite{[Boh58]} it has been largely
attributed to BCS pairing, but there are in fact a number of 
mechanisms that can contribute 
\cite{[Bri05],[Sat98w],[Rut98],[Rut99],[Ben00e],[Dob01],[Dug02a],[Dug02b]}.  
In this work we want to study the performance of BCS and its
Hartree-Fock-Bogoliubov (HFB) extension to the global body of
data, taking an energy density functional and pairing functionals 
that are in common use.  In that way, we hope to provide a 
benchmark to assess future improvements in the theory.  Since we do
not consider 
%
all
 mechanisms to generate the OES,
our conclusions must be tentative.

There are many DFT surveys that treat individual isotope
chains,
%
 e.g., \cite{[Dob01c],yu03}, with the $Z$=50 isotope chain a favorite for calculation
of pairing properties.  We shall see, however, that it can be quite
misleading to draw general conclusions without examining the whole 
body of OES data.  Also, in much of the literature
the OES was not obtained from differences of calculated
binding energy but rather inferred from the 
%
average HFB 
gap parameters, as, e.g., in 
Ref.~\cite{[Hil02]}.
We also mention the global mass tables 
by the Brussels-Montreal collaboration  \cite{[Gor02],[Gor07],[Cha08]}.  
While this work achieves a good performance on binding energies, it
deviates from the framework of DFT by adding phenomenological modifications
to the theory.  In particular, the pairing strength may depend on local
densities but it is hard to justify an explicit dependence on the number
parity as is assumed in ref. \cite{[Cha08]}.  

There are numerous measures of the OES
 in the literature, including 3-point,  4-point, and 
5-point difference formulas \cite{[Boh69],[Jen84],[Mad88],[Mol92],[Ben00e]}.  
In this work we will use the 3-point formula 
$\Delta^{(3)}_o$ as advocated in Ref.~\cite{[Sat98w]} and also used in 
Refs.~\cite{[Dob01],[Hil02]}.
For odd neutron number $N$, it is defined by the
binding energy difference
\be
\Delta^{(3)}_o(N) = {1\over 2} \left[ B(N+1)+B(N-1) - 2 B(N)\right].
\ee
In the 
following, we shall call this quantity the {\it neutron OES}. Our survey
will cover the proton OES as well.
One advantage of the \delo~statistic is  
that it can be applied to more experimental data than the higher-order
ones.
Another advantage is that it suppresses the smooth contributions
from the mean field to the gap.
The other
3-point indicator, $\Delta^{(3)}_e(N)$ with $N$-even, is less interesting
for our purposes because it is more sensitive to single-particle energies.

This paper is organized as follows. Section~\ref{Method} outlines
the theoretical DFT framework employed in this work.
In Sec.~\ref{Database} the selection of experimental data used in the survey
is discussed.
The results for selected spherical and deformed  isotopic/isotonic chains are presented in
Sec.~\ref{Chains} while the global performance of our pairing models
is analyzed in Sec.~\ref{Global}. Finally, Sec.~\ref{conclusions} contains the main
conclusions and perspectives.

\section{Methodologies}\label{Method}

We carry out two independent surveys with the same Skyrme functional  SLy4
\cite{[Cha98]}
in the particle-hole channel.
The pairing functional uses the zero-range
density-dependent $\delta$
 interaction:
\begin{equation}\label{deltap}
V(\mathbf{r},\mathbf{r'})=V_0\left(1-\eta \frac{\rho(\mathbf{r})}{\rho_0}\right)
\delta(\mathbf{r}-\mathbf{r'}).
\end{equation}
Here $V_0<0$ is the pairing strength,
$\rho(\mathbf{r})$ is the isoscalar nucleonic density,
and $\rho_0$=0.16\,fm$^{-3}$.  We have performed global calculations for 
$\eta$=0, 0.5, and 1,   called 
volume, mixed, and surface,
 pairing, respectively. The volume pairing interaction  acts
primarily inside the nuclear volume while the surface pairing
generates pairing fields peaked around or outside  the  nuclear
surface. As discussed in Ref.~\cite{[Dob01a]}, different assumptions about
the density dependence can result in notable differences of pairing fields
in neutron rich  nuclei.

The two surveys were carried out assuming two different theoretical frameworks
for the pairing channel, the BCS and the Hartree-Fock-Bogoliubov (HFB).
The details are described in the two subsections below.

\subsection{HF-BCS with {\ev}}

The HF-BCS extension of the nuclear DFT  can be
defined very concisely.  The ordinary variables in the theory, 
namely the orbital wave functions $\phi_i$ expressed in some basis, are 
augmented by the BCS amplitudes $v_i$.  
Specifically, one defines the
BCS $v_i$ and $u_i$ amplitudes  for each orbital and 
calculates the ordinary DFT energy from its functional using the 
density matrix $\rho({\bf r},{\bf r}') = \sum_i v^2_i \phi^*_i({\bf r})
\phi_i({\bf r}')$. 
To this is added the pairing energy functional, given by
\be
E_{pair} = \sum_{i\ne j} V_{ij} u_iv_i u_jv_j + \sum_i V_{ii} v_i^2\;\;
\label{eq:pair}    
\ee
where $V_{ij}$ are the matrix elements of the pairing interaction.

We use the  code {\ev} \cite{[Bon05]} to carry out the HF+BCS 
computations. 
 {\ev}  solves the HF+BCS equations for
Skyrme-type functionals via a discretization of the individual
wave-functions on a 3D Cartesian mesh and the imaginary time method.
(In {\ev}, the  pairing functional (\ref{eq:pair}) is
approximated as $\sum_{i,j} V_{ij}u_iv_iu_jv_j$.)
The pairing interaction matrix elements are those of
a  density-dependent contact interaction (\ref{deltap}).
 Contact interactions can only be used in truncated 
orbitals spaces; the calculations use the same truncation as is 
Ref.~\cite{[Ben06]}, namely an energy window of 10\,MeV around the Fermi level.

As we will see later, the OES only fluctuates about an average trend by
$\sim$0.3\,MeV, putting a  
high demand on the accuracy and the nucleus-to-nucleus consistency
of the self-consistent mean field calculations.  The usual iteration
procedure in {\ev}  appears to be adequate to achieve accuracy
at the 100 keV level in several hundred iterations at a 
fixed deformation. (Here accuracy
means with respect to the fully converged minimum of the numerically
implemented energy functional. This numerical  functional may contain approximations
that give a larger error with respect to the mathematically defined
functional.  In the case of {\ev}, the lattice representation of the kinetic
operator results in an error of the order of one MeV in heavy nuclei that varies
very smoothly with $A$.  Thus it largely cancels in the calculation of
\delo.)  Finding the minima irrespective of deformation is less
straightforward.  We adopted the following protocol to determine them.
We first build a table
of DFT energies and orbital wave functions of the relevant even-even
nuclei, using the minimum energy deformations from the table calculated
in Ref.~\cite{[Ben06]}.  The relative energies of spherical 
and deformed 
configurations are quite sensitive to the pairing interaction, so
in the cases where the spherical configuration in that table has an 
energy close to the deformed minimum, the spherical was also tested.
When it came out lower with the new pairing interaction, it replaced
the old entry in the new table.  Next, the table was refined in an
iterative way using only the deformation information about neighboring
even-even nuclei.  If two neighboring nuclei have substantially different 
deformations in the table, each
configuration must be tested in both nuclei.  If taking the lower
energy configuration results in a change, the process is repeated on
the neighbors of the replaced  nucleus.  This is continued until 
there are no further changes in the even-even DFT solutions.  

Once the DFT table of even-even nucleus is finalized, the odd-$A$ nuclei 
are calculated starting from  the DFT solutions
for the neighboring even-even nuclei.    
We performed the calculations using the so-called filling approximation
for the odd particle \cite{[Per08],[Ber08a]}.  The odd particle is assumed 
to occupy an orbital defined
by its position in the list of orbitals ordered by single-particle energy.
That orbital is blocked by setting $v^2=u^2=0.5$ in the calculation of
all ordinary densities, and omitting the orbital in the summation 
in Eq.~(\ref{eq:pair}).
During the self-consistency iterations,
the blocked orbital evolves along with the others, and thus may change
character if the relative ordering of the levels changes.  
Note that the filling approximation gives 
equal occupation numbers to both time-reversed partners, and therefore
misses the effects of time-odd fields on the OES.

Our protocol to find the most favorable orbital to block was to examine
the five orbitals around the Fermi level of the neighboring even-even
system.    We also tested configurations generated from the DFT solution
for the even-even nucleus with one more nucleon than the target odd-$A$
nucleus. Thus the total number of odd-$A$ configurations tested is 
ten:
five
starting from the lighter even-even core  and five starting from the
heavier one.

Since the objective is to determine the level of accuracy that can be
achieved, the calculations were carried for the nuclei in the data
set for a number of values of the pairing strength $V_0$.  The results
below are reported for a value $V_0$ close to that which minimizes the
average residual in the {\delo} data sets, taking neutron and protons 
independently.  

\subsection{HFB with  {\hfbtho}}

The HFB calculations were carried out with the axial 2D HFB solver
{\hfbtho} \cite{[Sto05]} 
that has recently been improved by implementing the
modified Broyden mixing \cite{[Bar08]}  to accelerate the convergence
rate.

The even-even nuclei are calculated first.  An initial set of configurations
is generated by performing constrained minimizations on a
quadrupole deformation mesh. Typically, there are 20 calculations having
deformations in the range $-0.5 < \beta < 0.5 $ with a mesh spacing
$\Delta \beta = 0.05$. Next, we turn off the constraint to find the local minima of the energy as a
function of $\beta$. When there are multiple minima, we select up to
three for further processing, taking no more than one of oblate and
prolate deformation, and also the spherical solution if it is a local
minimum. The final step is to perform unconstrained minimizations on the 
selected configurations. The iterative minimization is carried out until the
maximum change of the matrix elements of the HFB matrix elements falls
below $0.0001$\,MeV. However, in one case the iteration converges to  
a limit cycle with energies oscillating by 0.004\,MeV.  Since this is well
below the accuracy needed here, we accepted the (lowest) calculated
energy.   

The minimization for odd $N$ or $Z$ is started from the candidate
configurations produced at the second stage of the even-even
calculations. 
As in the BCS, the odd nucleus is treated in the 
filling approximation, by blocking one of the orbitals. 
Here one has to specify which orbitals to block to
generate the odd-nucleon configurations.  The blocking candidates are
determined by examining the HFB quasiparticle spectrum of the
neighboring even-even nucleus with smaller number of nucleus.  
Tested are all one-quasiparticle configurations with quasi-particle energies 
below the  energy cutoff   $E_{1qp,cut}$ which is  
not smaller than 2 MeV for heavy nuclei  and not bigger
than 8 MeV for very light systems.
For most nuclei, we take
$E_{1qp,cut}$=$25/ \sqrt{A}$~MeV.
 As in the last step for even-even nuclei, unconstrained
calculations are performed for all candidate configurations to find the
absolute minimum energy.

In the second variant of HFB calculations (HFB+LN) we performed an
approximate particle number projection (before variation) using the
Lipkin-Nogami (LN) method \cite{[Lip60],[Nog64]}. The practical
implementation of the LN treatment follows Ref.~\cite{[Sto07]} where the
method was compared to the full particle number projection.

In HFB and HFB+LN calculations we employed the orbital space extending
to 20 major harmonic oscillator shells. For the pairing interaction, there is no lower
energy orbital cutoff and for the upper equivalent energy cutoff  we
adopted the commonly used value  of 60 MeV \cite{[Dob04a]}. The
calculations were first performed with a standard pairing strength 
$V_0^{\rm std}$ adjusted to the average pairing gap in 
$^{120}$Sn according to the procedure of \cite{do95}.  The values obtained are $V_0^{\rm std}=-258.2$ and $-284.57$
for the HFB and the HFB+LN calculations.   However, following our first 
survey, we found these strengths to be
too small to make a good global fit to OES.  We
then increased the pairing strength by a factor of 1.2 and recalculated the mass
table from scratch.   Both sets of tables are available through the UNEDF
SciDAC collaboration \cite{UNEDF}. Our fit to the global systematics is then made
using a linear fit of the data sets of 
the two mass tables:
\begin{equation}\label{inter}
M(x) = x M(V_0^{\rm std}) + (1-x) M(1.2\cdot V_0^{\rm std}).
\end{equation}
The  effective pairing strengths obtained in this way is  given by
\begin{equation}\label{VHFB}
V_0^{\rm eff} = (1.2-0.2x) V_0^{\rm std}.
\end{equation}
The values of $x$ and the derived pairing strengths are reported in Table \ref{pair}.
\begin{table}
\caption{
Effective strengths for the pairing interactions 
derived in the present study by means of Eq.~(\ref{inter}).\label{pair}.
Units of $V_0$ are MeV-fm$^3$.  Also the Values of the fit parameter $x$ 
are given in paratheses.
}
\begin{ruledtabular}
\begin{tabular}{lcrrr}
theory  &   density dependence  & $V_0^{nn}(x)$ & $V_0^{pp}(x)$ \\
\hline
BCS     &   volume & 465.0   & 490.0  \\          
        &   mixed & 700.0   &  755.0  \\
        &  surface & 1300.0   & 1462.0   \\       
HFB     &  mixed      &318.1(0.41)    & 352.0(-0.18)   \\    
HFB+LN  &  mixed     &300.5(0.18) & 332.6(-0.44)   \\       
\end{tabular}
\end{ruledtabular}
\end{table}

\subsection{Other methodological aspects}

In setting up the calculational protocols for the survey, we of course
scrutinized cases where the residuals between theory and experiment were
large.  The residuals are very sensitive to numerical inaccuracies, and
their detailed analysis often brought to attention problems with
calculated numbers due to, e.g., incorrectly assigned configurations or
the lack of convergence in self-consistent iterations.  We could then
refine the protocols by making a broader screen or demanding higher
precision to produce more accurate tables.  From this standpoint we
found {\delo} a very useful indicator. Also, the fact that the theory is
variational is a tremendous help: any change in protocol that gives
lower energies is necessarily an improvement.

Apart from the numerics, it should be noted that the pairing gap is a very 
strong function of the pairing strength.  For example, for the HFB
calculations, we found that increasing the pairing strength by 20\% 
from the value fitted to
 $^{120}$Sn, the average neutron pairing gap
increases  by a factor of 2.3.  
This sensitivity is not surprising.  
A typical nucleus in our set  has some deformation
and does not have a large single-particle degeneracy at the Fermi level. 
This implies that 
its BCS condensate is weak.
Under these
conditions, the BCS gap increases very quickly from a zero value at some
finite strength of the interaction.   In fact, we find that roughly 20-30\%
of the calculated OES contain a nucleus whose BCS or HFB condensate  
has collapsed.  Even in infinite systems having
no single-particle gap at the Fermi energy, the condensate $\Delta$ grows exponentially
with interaction strength in the well-known BCS formula.

\begin{figure}[htb]
\centerline{\includegraphics[trim=0cm 0cm 0cm
0cm,width=0.5\textwidth,clip]{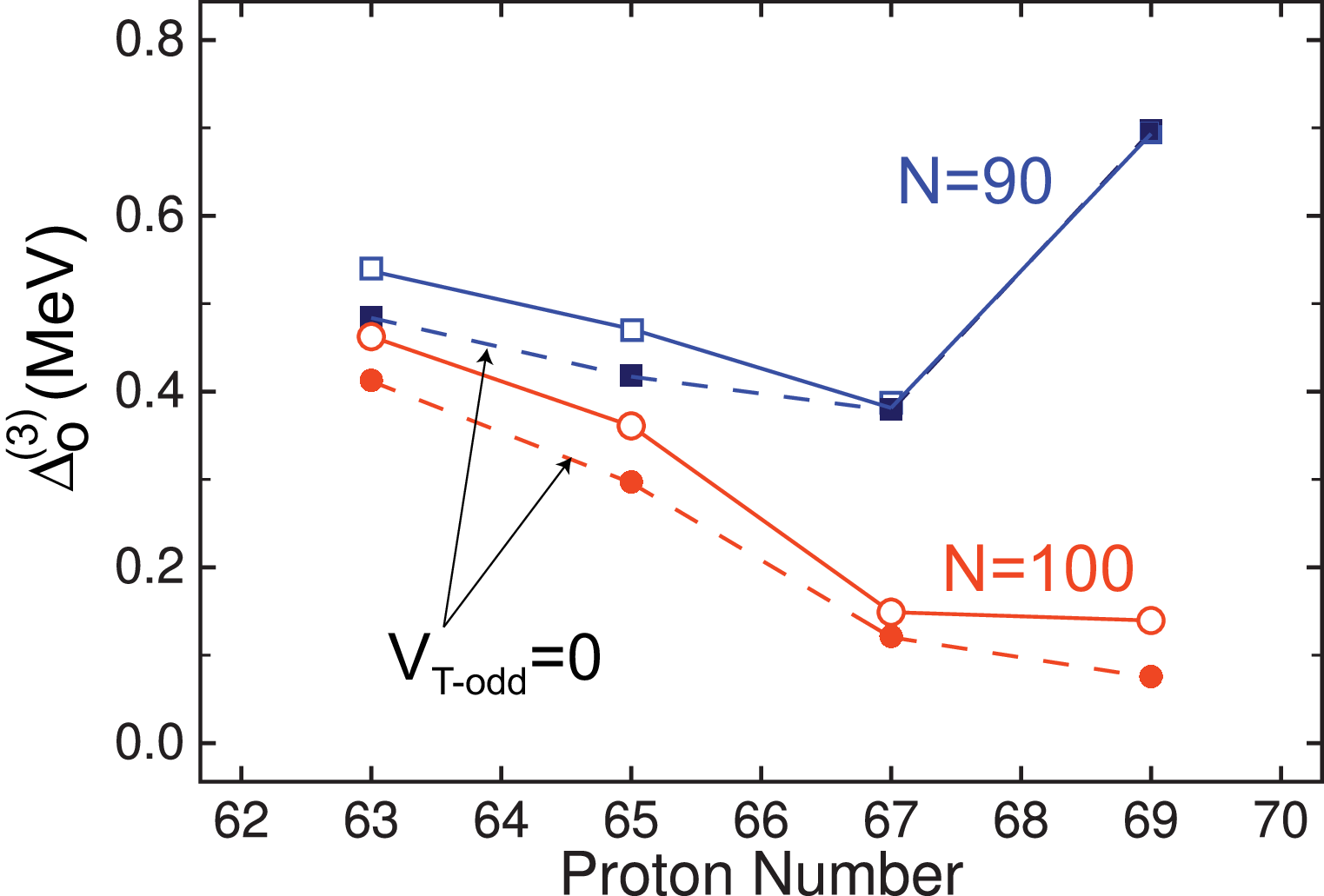}}
\caption{(Color online)  Impact of time-odd terms on the proton OES
{\delo}
in the $N$=90 and 100  isotonic
chains calculated in the HFB theory with SLy4 functional and mixed pairing.
Solid lines connect the values of proton OES including time-odd fields;
dashed lines show the values of OES  omitting the time-odd fields.
Calculations were  performed with the code {\sc hfodd} of Ref.
\cite{[Dob05c]} in a deformed HO basis containing 680
deformed harmonic oscillator states.
}
\label{TODD}
\end{figure}

An exact implementation of HFB requires the breaking of the
time-reversal symmetry in the intrinsic frame of an odd-$A$ nucleus. The 
resulting time-odd mean fields  contribute to the binding energy
of the odd-$A$ system; hence,  the OES {\delo}  depends
on whether these time-odd terms are included or not
\cite{[Rut98],[Rut99],[Dug02a],[Dug02b],[Zal08]}. In general,  
time-odd mean fields are known poorly
and their effects differ from model to model. For instance, in the Skyrme
DFT calculations of Ref.~\cite{[Zal08]}, 
the time-odd polarizations systematically shift
the hole states down and particle states up in energy, while a different
result has been  obtained in the relativistic mean field approach \cite{[Rut98],[Rut99]}.
For the purpose of  
illustration, Fig.~\ref{TODD} shows the OES calculated
in the HFB for isotonic chains including the effects of time-odd fields.  
A detailed analysis of
resulting polarizations will be published elsewhere \cite{[Sch08]}.
It appears that
the contribution is less that $100$ keV in heavy elements such as the rare
earths and actinides, but larger values have been found
with other models, eg. \cite{[Rut99],[Dug02b]} and in light nuclei. In any 
case, 
the density functional has not been fitted to time-odd properties, so no
quantitative evaluation of their effect is possible.
In this work, we employ the filling approximation, neglecting all time-odd
interactions.
%

\section{Experimental data base }\label{Database}

The data base for our survey derives from the 2003 atomic mass evaluation
 \cite{[Aud03]}.  The accuracy requirements our purposes is
of the order of 100 keV, so we only use the masses whose evaluated
experimental errors were less than 200 keV.  This gives  908 
mass triplets
for the neutron OES and 864 for protons. However, we made 
additional cuts to remove nuclei for which some physics is obviously
missing from our BCS or HFB theory.  First of all, in odd-odd 
nuclei there is an additional neutron-proton pairing effect.  Its origin 
is not completely clear \cite{[Fri07]}, but it is obviously beyond the scope of the 
pairing theory we use here.  We therefore do not include binding energy
triplets containing odd-odd nuclei.  Another phenomenon in nuclear
binding that affects the OES is the so-called Wigner energy, an increased
binding at $N$=$Z$.  This might also be a neutron-proton pairing
effect  (see, e.g., Refs. \cite{[Sat97],[Sat01]}). We therefore eliminate triplets that contain
$N$=$Z$ nuclei.  More generally, mean-field approximations becomes doubtful
when the number of particles is small.  Some restriction of light nuclei
is imposed by a cut on
particle numbers, requiring that the neutron and proton numbers be greater
than 8, corresponding to nuclei heavier than $^{16}$O. Finally, we only include nuclei with
$N$$>$$Z$. There are only a few OES on the proton-rich side satisfying our
other criteria, and keeping a fixed sign of the isospin will permit us 
to make some qualitative statement about the isospin dependence of the
interaction.  With all these cuts, there
are left 443 triplets for the neutron OES and 418 for the proton in
our final data set.

\begin{figure*}

\includegraphics [height = 5.7cm,viewport=00 11 338 240,clip]{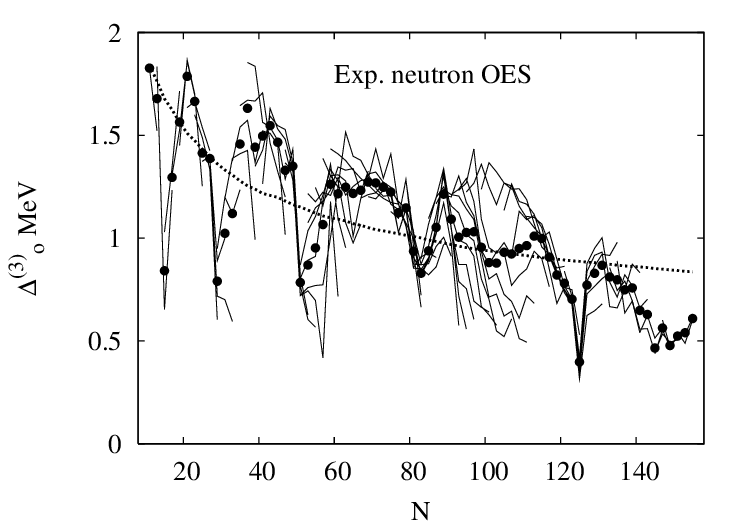}%
\includegraphics [height= 5.7cm,viewport= 61 11 399 240,clip]{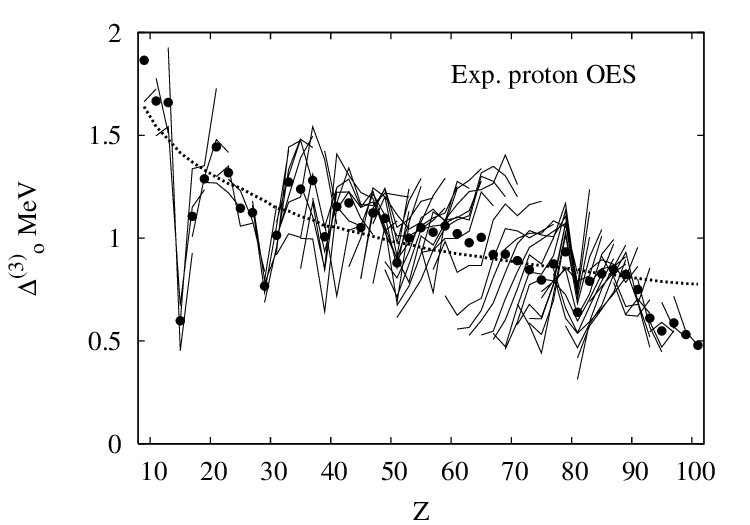}
\includegraphics [height = 5.7cm,viewport=00 11 338 240,clip]{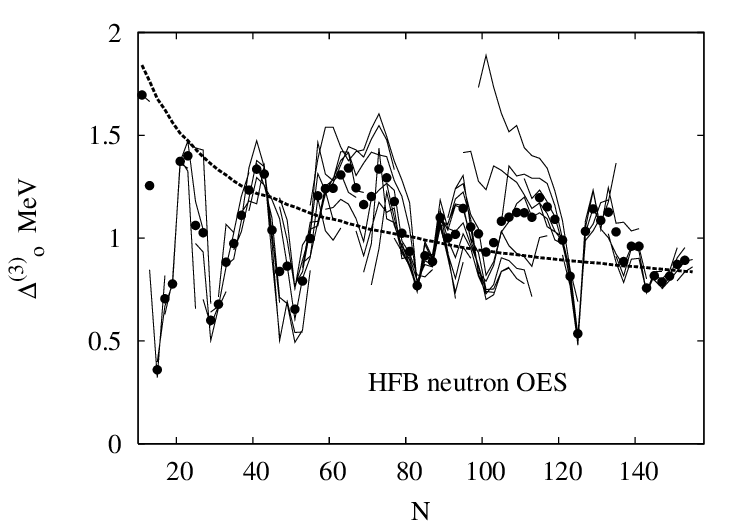}%
\includegraphics [height= 5.7cm,viewport= 61 11 399 240,clip]{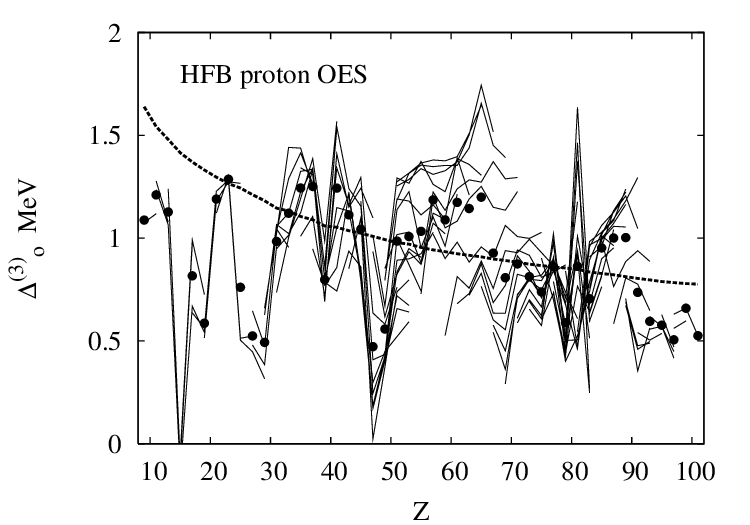}
\includegraphics [height = 5.7cm,viewport=00 11 338 240,clip]{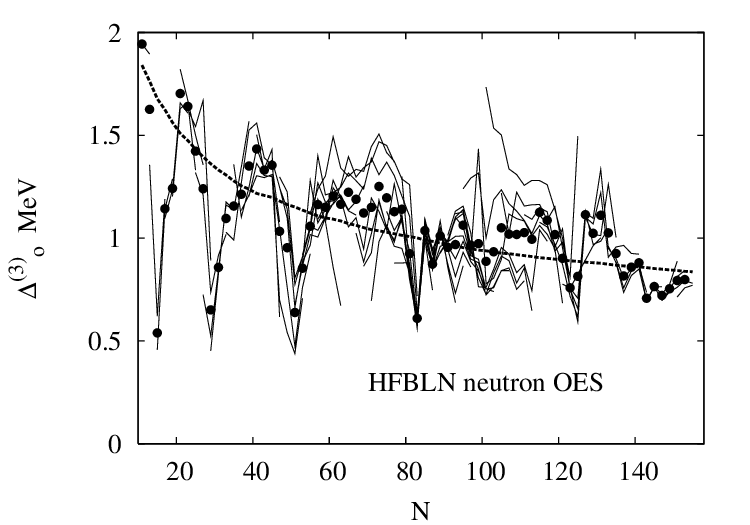}%
\includegraphics [height= 5.7cm,viewport= 61 11 399 240,clip]{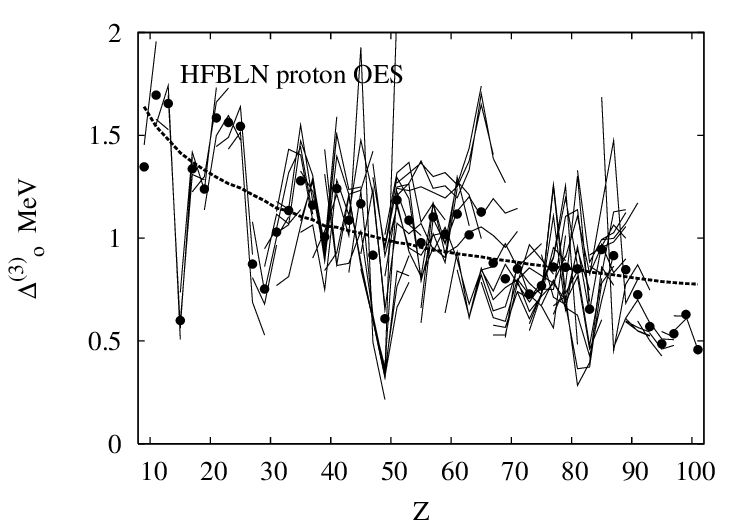}
\includegraphics [height = 6.0cm,viewport=00 00 338 240,clip]{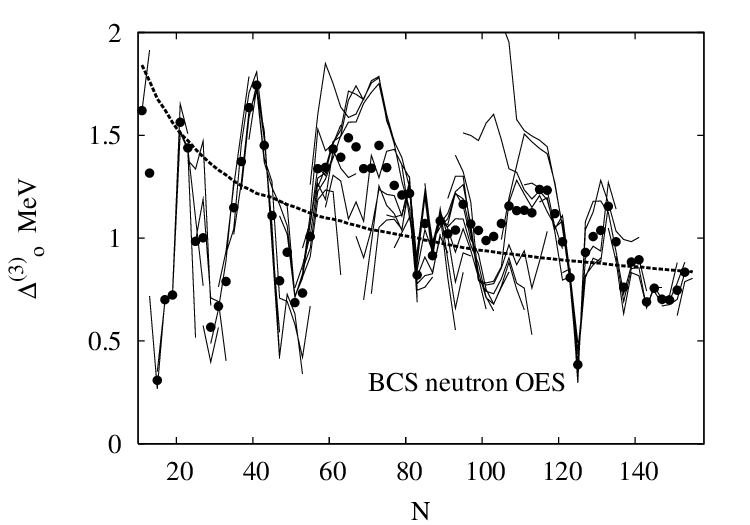}%
\includegraphics [height= 6.0cm,viewport= 61 00 399 240,clip]{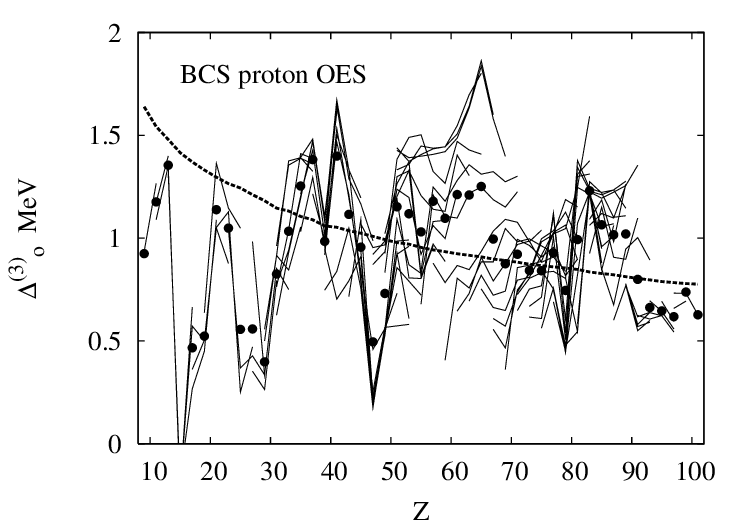}

\caption{\label{fi:all} OES as a function of
nucleon number.  From top to bottom, panels show {\delo} for
experiment, HFB, HFB+LN, and BCS treatments, respectively.  Circles show
values obtained by averaging over nucleon number of the opposite isospin
to that of the OES.  The calculation used the SLy4 Skyrme energy functional
and a pairing interaction with the mixed density dependence.
}
%
%
\end{figure*}

The sets of neutron and proton {\delo} are plotted as a function of
neutron and proton number in the top panels of 
Fig.~\ref{fi:all}.  Lines connect the values of OES  
for the same number of nucleons of
the opposite kind.  It is common to plot OES as a function of $A$,
but plotting it with respect to nucleons of the same kind makes clearer
the origin of fluctuations.  Probably the largest cause of fluctuation 
is the variation in single-particle level densities and the character of
the level at the Fermi energy.  This obviously depends strongly on the
number of nucleons of the same kind, and this motivates the choice of 
abscissa variable.  The single-particle level densities may also change
with the different numbers of nucleons of the other kind, particular
if the additional nucleons causes a large change in deformation.  Such
effects should be visible in the variation of OES at fixed value of
the abscissa.   To emphasize the like-nucleon fluctuations, we also show
as solid circles the average with respect to
nucleons of the opposite kind.  

One can see that the shell effects are large and there are large fluctuations on the
scale of major shell spacings. For the neutron values of {\delo},
 there are strong dips
in the $Z$-averaged values at $N$=15, 29, 51, 83, and 125,
i.e., 
in the vicinity of  shell closures.  We shall call this phenomenon 
{\it gap quenching}.   Obviously, the OES is reduced when one of the
three nuclei is at a magic number where the gap in single-particle
energies is large.  We will examine the effect in more detail below,
in presenting the theoretical OES. In Table~\ref{exptable} we show the extreme OES cases--
either the largest or the smallest in our experimental data sets. 
\begin{table}[htb]
\caption{\label{exptable}
Nuclei from the data base adopted for our survey
with the largest and smallest experimental values of {\delo} (in MeV).
}
\begin{ruledtabular}
\begin{tabular}{lcccc}
   & \multicolumn{2}{c}{largest}  &\multicolumn{2}{c}{ smallest}  \\
   & $(N,Z)$ & \delo & $(N,Z)$ & {\delo} \\
\hline
neutrons  & (21,16) &1.87 &  (125,82)& 0.32\\
protons & (12,9)& 2.07   & (126,81)& 0.31 \\
\end{tabular}
\end{ruledtabular}
\end{table}

Another observation that can be made about the neutron
OES is that the variations with respect to $Z$ are 
particularly large in
the regions $N$=50-60 and 95-110.  For the protons,
the regions of strongest
$N$-dependence are $N$=35-40 and 60-70. We will see that this
is associated, at least in part, with changing deformations.

The averages and variances for neutron and proton OES {\delo} in
the data set are 1.04$\pm$0.31 and 0.96$\pm$0.27\,MeV, respectively.  
The lower average for the protons can be largely attributed to
the Coulomb interaction: in the liquid drop formula,
the term $a_c Z^2/A^{1/3}$ gives an average value of 0.11\,MeV for the
proton {\delo} in the data set.  
There is some overall dependence of the OES on mass number $A$ which may be seen by 
visual inspection of Fig.~\ref{fi:all}.
For more discussion of the global mass dependence
of the OES, we refer the reader
to  Refs.~\cite{[Mol92],[Sat98w],[Hil02]}. 
The smooth $A$-dependence seems rather weak compared
to the local nucleus-to-nucleus fluctuations caused by shell effects, 
but parameterizing it in some
way can give improved fits.  For example,
the phenomenological parametrization \cite{[Boh69],[Olo08]}
\begin{equation}\label{Dave}
\tilde{\Delta}=\frac{c}{A^\alpha}
\end{equation}
with $c$=4.66\,MeV (4.31\,MeV)
and $\alpha$=0.31
gives an rms residual of 0.25\,MeV on the neutron (proton) data set. 
The global trends given by Eq.~(\ref{Dave}) are shown as the dashed lines in
 Fig.~\ref{fi:all}.  For the 
sake of the  plot,
 we averaged over nucleon number of the
opposite kind just as was done to produce experimental averages.

An important question about the global systematics is whether there is an
isospin dependence of the pairing interaction.  It is clear from
Fig.~\ref{fi:all} that there can be strong interaction between the pairing of
one kind on the numbers of the other kind, but as mentioned it could be
due to other effects such as shape changes. The isospin dependence has been 
examined in Ref.
\cite{[Mol92]} but no significant effect has been found  (see also
Refs.~\cite{[Vog84],[Jen86],[Sta92]} for more discussion of isovector
trends). On the other hand, a recent study \cite{[Lit05]} of the
OES of nuclear masses for isotopic chains between the
proton shell closures at $Z$=50 and $Z$=82, including nuclei with
extreme isospins, has claimed a significant isospin dependence of
pairing.   We will discuss in more detail the possible evidence for an
isovector dependence of the interaction in Sec. \ref{isospin}.

\section{Results: local comparisons}\label{Chains}

\begin{figure}[htb]
\centerline{\includegraphics[trim=0cm 0cm 0cm 0cm, 
width=0.4\textwidth, clip]{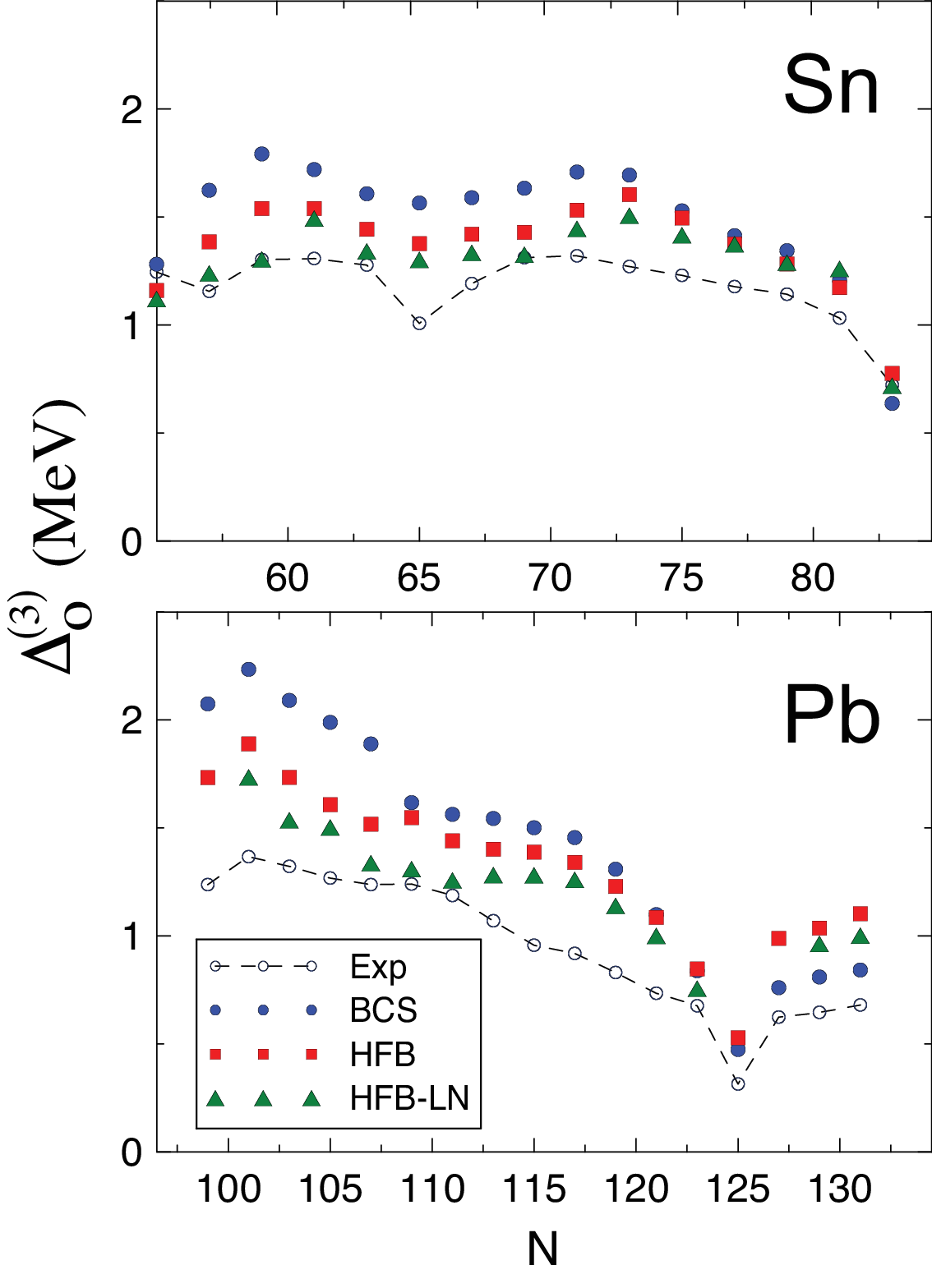}}
\caption{\label{sphgaps} (Color online)
Calculated (HF+BCS, HFB, and HFB+LN) and experimental values of {\delo} for neutrons
in  the Sn (top) and Pb (bottom) semi-magic 
isotopic chains. In all calculations, the pairing interaction 
was taken in the mixed pairing form
($\eta$=0.5)  with strength $V_0$ (or $x$ in Eq.~(\ref{inter})) adjusted to
global systematics.	
}
\end{figure}

\begin{figure}[htb]
\includegraphics[width=0.4\textwidth]{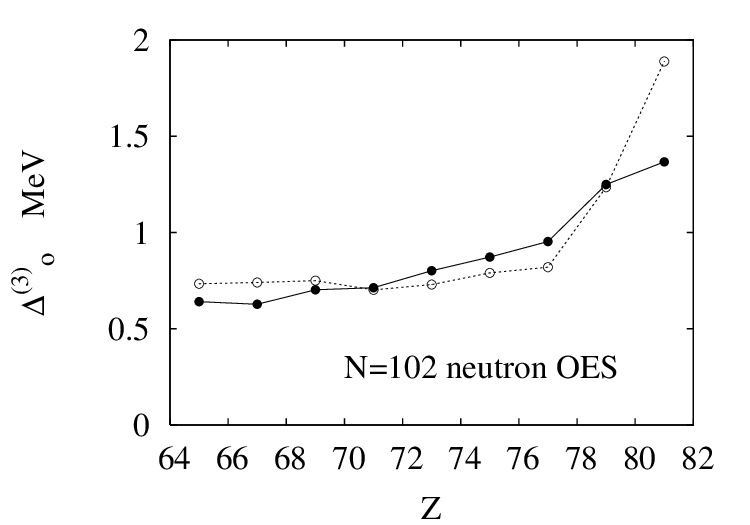}
\caption{\label{at102} 
Experimental (filled circles) and HFB (open circles) neutron OES for  $N=102$
as a function of $Z$. The chain starts in the  well-deformed
lanthanides and ends next to the singly magic $^{184}$Pb. 
}
\end{figure}

We begin our comparison between theory and experiment with 
two spherical semi-magic
isotope
chains, Sn and Pb.  The
results of the calculations are shown in Fig.~\ref{sphgaps}.
%
For all three treatments of pairing, the  trends of predicted OES 
are consistent with the data, concerning both global and local variations.
In particular, theory reproduces the flatness in the Sn isotopes up
to the quenched gap at $N=83$, and the 
downsloping trend from the light Pb
isotopes up to the quenched gap at $^{207}$Pb.  In the Sn isotope chain,
there is a small dip at $N=65$ which might be attributed to a neutron
subshell closure at $N=64$. In any case, the theories all predict
a shallow local minimum. 
 When confronted with experiment, HFB appears
to do better slightly than HF+BCS.
As mentioned earlier, the strength of the pairing interaction was fit 
to the overall systematics, giving a somewhat too high an average OES 
in both spherical chains. 
%
\begin{figure}[htb]
\includegraphics[width=0.4\textwidth]{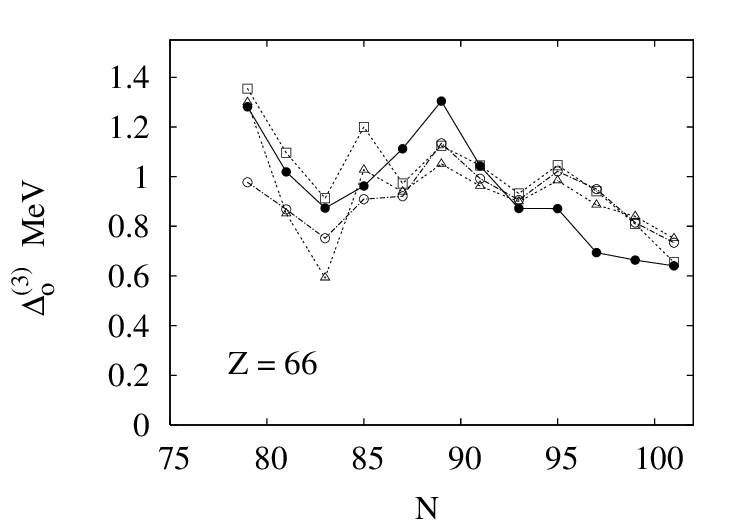}
\includegraphics[width=0.4\textwidth]{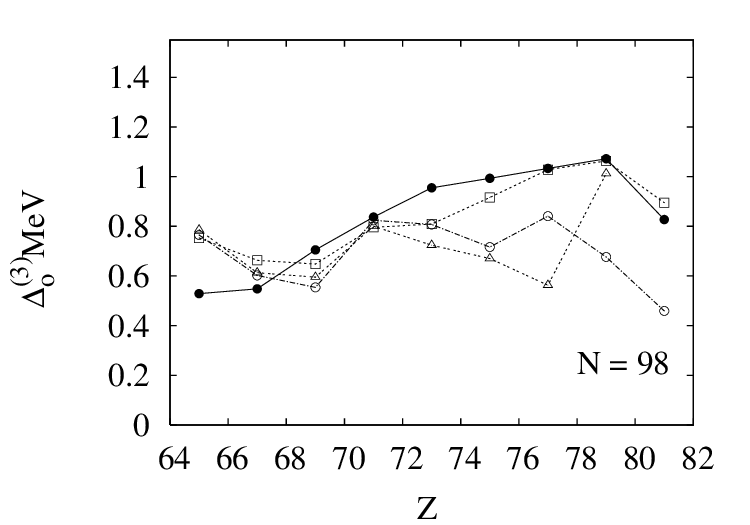}
\caption{\label{defgaps} (Color online)
Calculated and experimental values of {\delo} for neutrons (top, $Z$=66) and protons
(bottom, $N$=98) in rare earth nuclei, including strongly deformed systems.	
Filled circles: experiment; squares: BCS; open circles HFB; triangles:
HFB+LN.
}
\end{figure}
The performance of the theories for long isotonic and isotopic chains that
include deformed nuclei is  illustrated in Fig.~\ref{defgaps} and
\ref{at102}.  Fig.~\ref{defgaps} shows  
the neutron OES in the Dy isotopes ($Z$=66) and proton OES
in the $N$=98 isotones. Like in the  case of semi-magic nuclei, the agreement with experiment
is  good, in particular for well deformed nuclei where the mean 
field changes smoothly with
particle number. The effect of changing deformation is illustrated
by the region from
$A\sim160 $ to $\sim 190$ which starts deformed and becomes spherical
as $Z$ is increased from $66$ to 82.  The 
neutron values of OES for $N=102$ covering this transition 
region are shown in 
Fig.~\ref{at102}
as a function of $Z$.  The squares show the experimental OES,
which increase from about 0.6-0.8 MeV
for the lower $Z$ nuclei and goes up to $\sim 1.3$ MeV for the singly magic
$Z=82$ isotope.  The circles show the corresponding 
calculated HFB values of OES. The trend is very similar, but the theoretical
rise to $Z$=82 is 
sharper and higher than is seen experimentally.  Very likely, the increase
in single-particle level density going from deformed to spherical nuclei
is responsible for increasing trend in the OES.
%

The fact that our calculations overestimate OES  in
spherical nuclei may be partly attributed to the particle number
fluctuations. The pairing gap exponentially depends on the inverse of
the single-particle level density at the Fermi level, which is large in
spherical open-shell nuclei due to the $2j$+1 degeneracy (the limit of
pairing rotation). In deformed systems, the level density is reduced due
to the  Jahn-Teller effect \cite{[Rei84],[Naz94a]} and this gives rise
to the  overall pairing reduction; hence,  a transition towards the
transitional pairing  regime, in which the particle number fluctuations
are more important. Since the original pairing strength $V_0^{\rm std}$
was adjusted to the global data set containing far more non-spherical
nuclei than semi-magic systems, the resulting  deformation bias results
in too strong pairing correlations predicted for spherical nuclei, as
seen in Fig.~\ref{sphgaps}. This can be partly cured by considering
particle-number fluctuations. Indeed, as seen in Figs.~\ref{sphgaps} and
\ref{defgaps} the LN procedure slightly improves agreement with
experiment for spherical nuclei while still reproducing data for
deformed chains. 

\section{Results: global }\label{Global}
\subsection{Overview and shell effects}
The lower three sets of panels in Fig.~\ref{fi:all}
show distributions of neutron and proton OES for our three 
theoretical treatments: HFB, HFB+LN and  BCS.
To make the overall trends with
respect to shell filling more apparent, we also show the values
of the OES obtained by averaging over nucleons of the opposite
isospin.  Qualitatively, the three methodologies give rather similar
results.  
In all cases,
the trends of the predicted gaps are consistent with the data.  
The strong neutron gap quenching seen in the experimental OES at
the numbers $N=83$ and $N=125$ is reproduced in all three theoretical
treatments.    The gap quenching
is obviously dependent on the presence of shell closures, but the fact
that it does not invariably occur on both sides of the magic numbers
indicates that particular orbital properties must play a role.  
Since the HFB calculations
were performed in an axially symmetric basis, we can examine the quantum
numbers of the blocked orbital. 
In Table III
we show the quasiparticle orbital characteristics for odd nuclei
exhibiting quenched gaps.  The large quenching 
at $N=125$ can be understood as a spherical shell effect associated with the
$p_{1/2}$ shell at the Fermi level.  A $j=1/2$ shell will have
reduced pairing for two reasons.  First, there is no degeneracy within
the shell to correlate pairs, so all of the pairing has to come from
off-diagonal interactions to other shell.  Second, these couplings are
reduced because the spatial overlap of high-$l$ and low-$l$ orbitals is
poor.  Similar considerations apply to $N$=15 and $Z=81$, where the relevant
spherical shell is $s_{1/2}$.

While the location of the gap quenching is well reproduced by theory,
the magnitude of the effect is often exaggerated.  Most notably,
all the
experimental values of OES are positive, the
theoretical ones at $(N,Z)=(20-24,15)$ even have a
negative sign.

\begin{table}
\caption{\label{Kp} Characteristics of nuclei with quenched gaps in the HFB
calculations.  The listed nucleus is the one with smallest OES
at the given gap.  The calculated deformation $\beta$ is given in the
third column.  The last columns give the quasiparticle quantum numbers
angular momentum and parity $J^\pi$ for spherical nuclei and 
azimuthal angular momentum $K$ and parity $K\pi$ for deformed nuclei.
}
\begin{tabular}{|lccc|}
\hline
$N$ gap &    Nucleus &   $\beta$ & $j^\pi$ or  $K^\pi$ \\
\hline
15     &           $^{25}$Ne  &          0.00  &      $1/2^+$ \\
29     &           $^{52}$Ti &          0.08  &     $1/2^-$  \\
47-51  &           $^{87}$Kr   &         -0.09  &     $5/2^+$  \\
83     &           $^{147}$Gd   &       -0.03&     $7/2^-$   \\
125    &           $^{125}$Pb    &          0.00&     $1/2^-$\\
\hline
$Z$ gap &&& \\
\hline     
15     &           $^{39}$P  &           0.21  &     $1/2^+$ \\
29     &           $^{61}$Cu  &           0.10 &     $3/2^-$ \\
47-49  &           $^{111}$Ag   &          -0.22 &     $1/2^+$ \\
69     &           $^{173}$Tm &           0.33 &     $7/2^-$ \\
81     &           $^{203}$Tl  &           0.01&     $1/2^+$ \\
\hline
\end{tabular}
\end{table}
Comparing the different treatments of pairing, we see that
the fluctuations seen are
at the
same positions in BCS and HFB+LN as in the HFB.  However, the amplitudes of the fluctuations
seem somewhat larger in the BCS treatment but somewhat smaller in 
HFB+LN.  It is not clear why the BCS should emphasize the fluctuations,
but the fact that they are damped in HFB+LN is not surprising.
In both BCS and HFB the static pairing sometimes collapses in a significant
fraction of nuclei, as mentioned earlier. The pairing never collapses in the HFB+LN
treatment, so the OES should be smoother as one passes into a region
of weak pairing.

Large variations
with proton number are found around $N$$\sim$70 and in the region
$N$=100-120.  As mentioned earlier, the latter region is a transition
region between spherical and deformed ground states, and that will
affect the OES.  The HFB theory has a spike as well as quenched-gap
behavior at $Z=81$ in the proton gap.  
The origin of the spike appears to be the
coexistence of spherical and deformed configurations in light isotopes
of $Z$=81 and 82 nuclides.  At the phase transition point, there can be
a large static polarization contribution to the OES.

\begin{figure}[htb]
\centerline{\includegraphics
[trim=0cm 0cm 0cm 0cm, width=0.4\textwidth, clip]{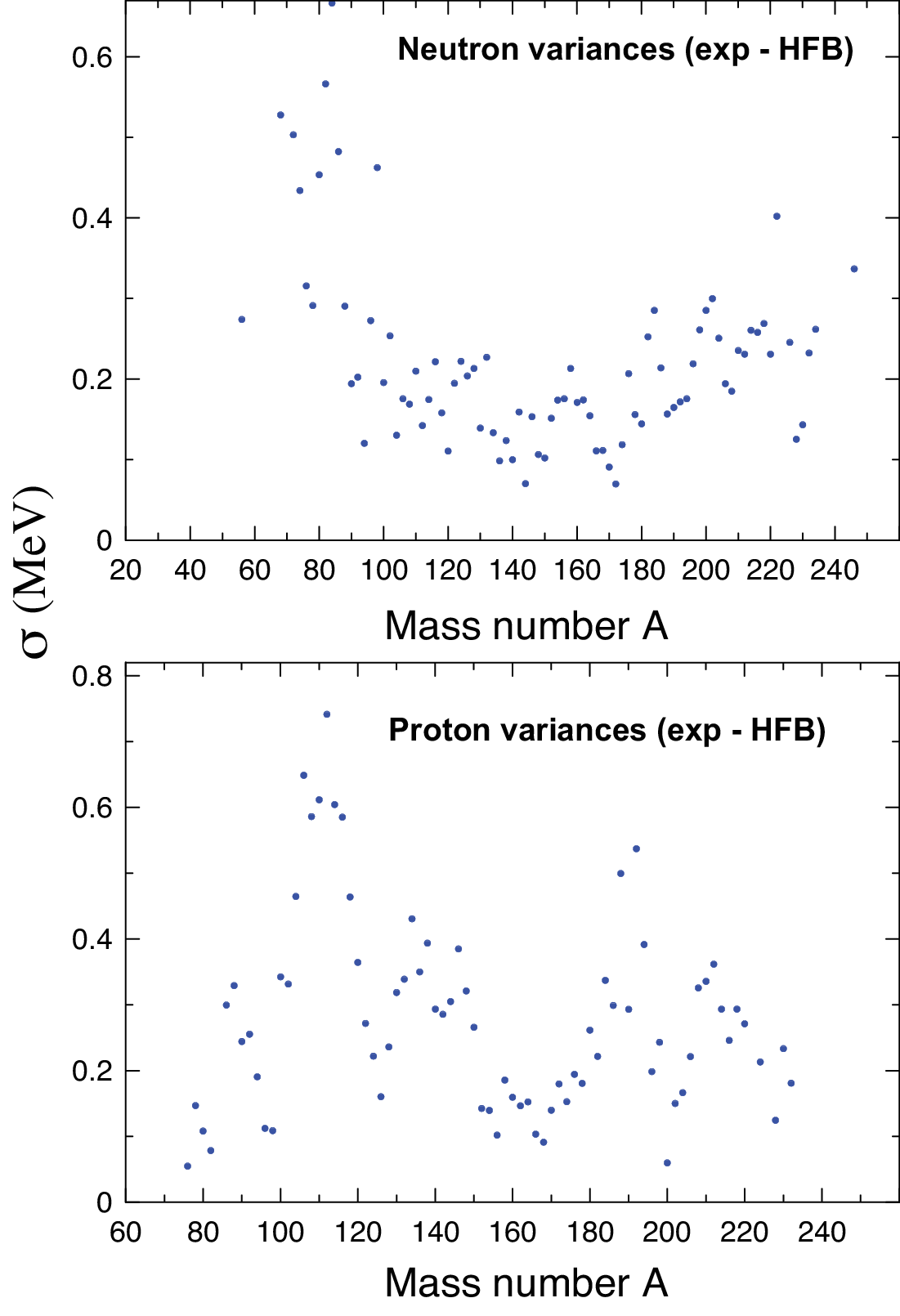}}
\caption{\label{variances} (Color online)
Neutron (top) and proton (bottom) average variances $\sigma$ between HFB and experimental
 {\delo} as a function of $A$. At each $A$, the variance $\sigma$ was obtained by
 averaging the  residuals over all  isobars available.
}
\end{figure}

To see where theory performs best, and where important physics is missing,
in Fig.~\ref{variances} we show the isospin-averaged residuals for our HFB
model. As expected, the best agreement is obtained for well deformed rare
earths and actinides whose properties vary smoothly with particle number.
The largest deviations are seen around shell closures and in the regions of
shape coexistence ($A$$\sim$90 for neutrons and $A$=110 and 190 for protons)
where dynamic shape fluctuations are known to strongly impact 
masses \cite{[Ben05a],[Ben06]}.
%

\begin{figure}[htb]
\centerline{\includegraphics
[trim=0cm 0cm 0cm 0cm, width=0.4\textwidth, clip]{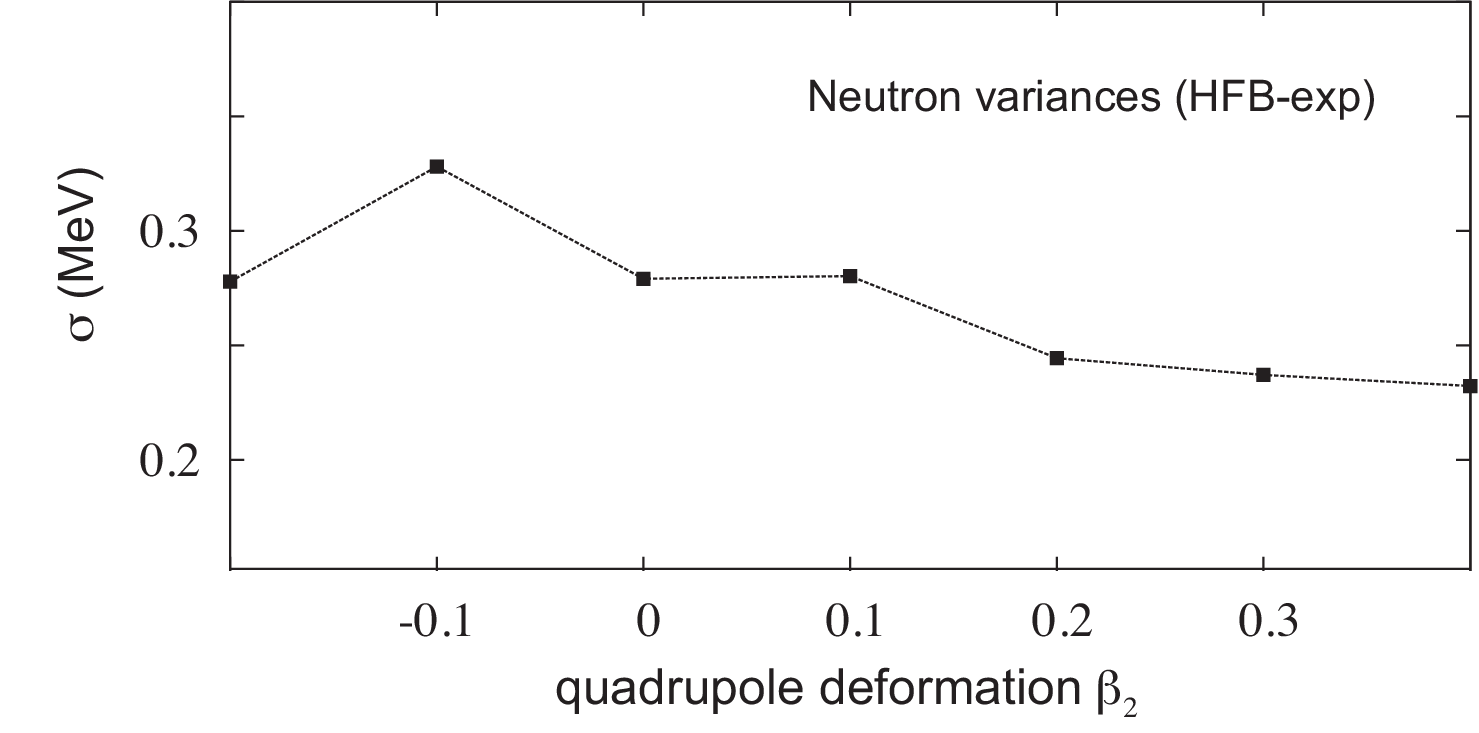}}
\caption{\label{deldef} 
Neutron average variances $\sigma$ between HFB and experimental
 {\delo} as a function of the  calculated quadrupole 
 deformation $\beta_2$. The deformation range --0.2$\le$$\beta_2$$\le$0.4
 was divided into bins with $\Delta\beta_2$=0.1, and each bin 
the variance $\sigma$ was obtained by
 averaging the  residuals over all nuclei available.
}
\end{figure}

To analyze more quantitatively the improvement in accuracy for deformed 
nuclei, we have examined the rms residuals for the neutron OES separating
the data into bins by the calculated values of the deformation $\beta$.
Figure~\ref{deldef} shows the averaged HFB residuals. As expected,
the residuals gradually decrease with deformation. The transitional/coexisting
nuclei with weakly-oblate shapes show the largest deviations from the data.
%

\begin{figure}[htb]
\centerline{\includegraphics
[trim=0cm 0cm 0cm 0cm, width=0.45\textwidth, clip]{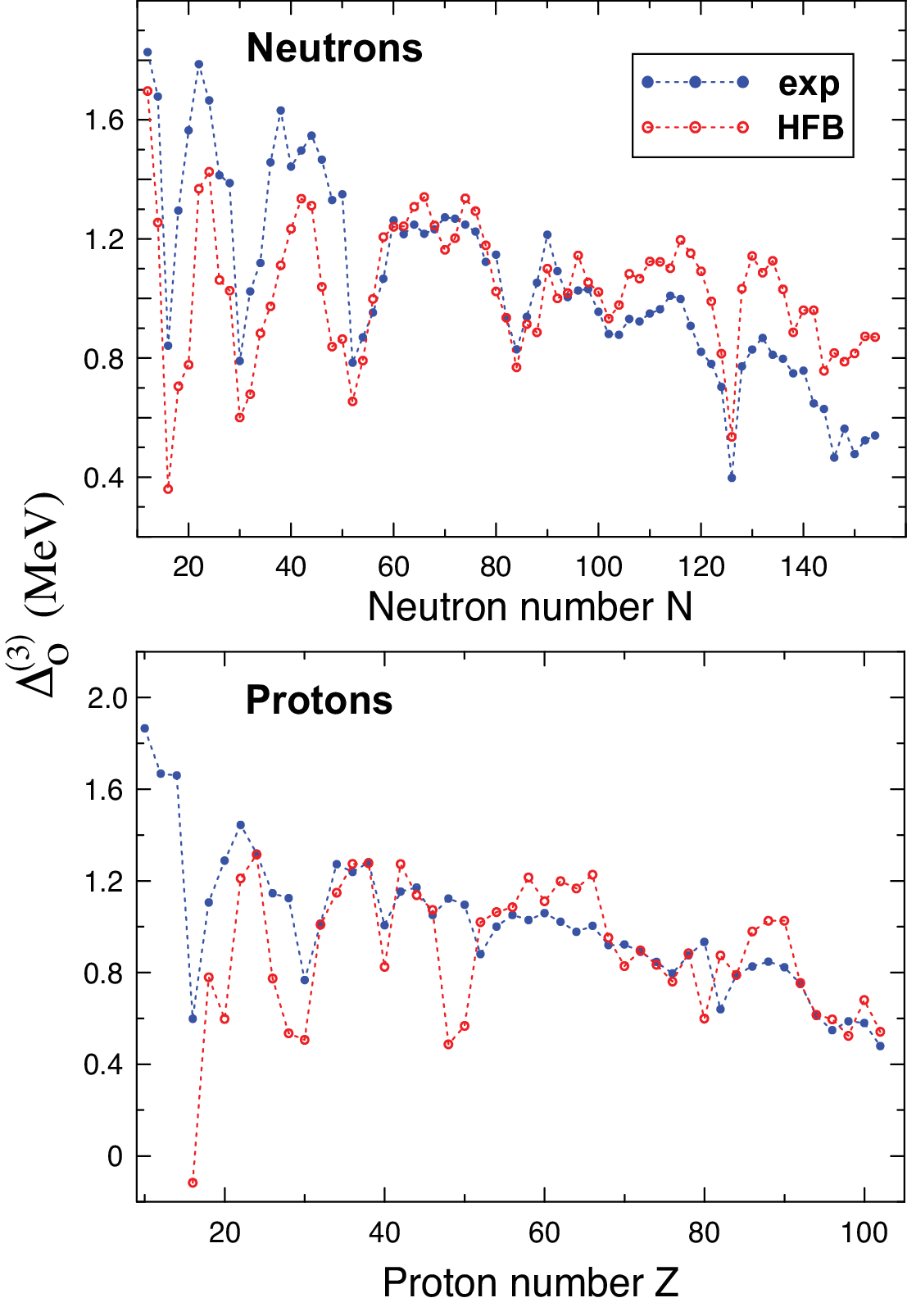}}
\caption{\label{fi:ave} (Color online) 
Comparison between calculated (HFB with mixed pairing) and experimental {\delo} values.
Top: $Z$-averaged values for neutrons. Bottom: $N$-averaged values for protons.
}
\end{figure}

The fluctuations in OES are obviously suppressed when one averages
over nucleons of the opposite isospin.  With that averaging, the comparison
between theory and experiment looks much better.  Fig.~\ref{fi:ave} shows
the theory-experiment comparison for the HFB methodology.  Besides seeing
the shell effects discussed above, one also sees better how the theory
performs with respect to the $A$-dependence of the pairing.  The theoretical
proton OES has an overall $A$-dependence that seems to accord well 
with the experimental trend.  For the neutron OES, however, the theory
is flatter than the experimental trend.  

\begin{figure}[htb]
\centering
\includegraphics [width=0.45\textwidth]{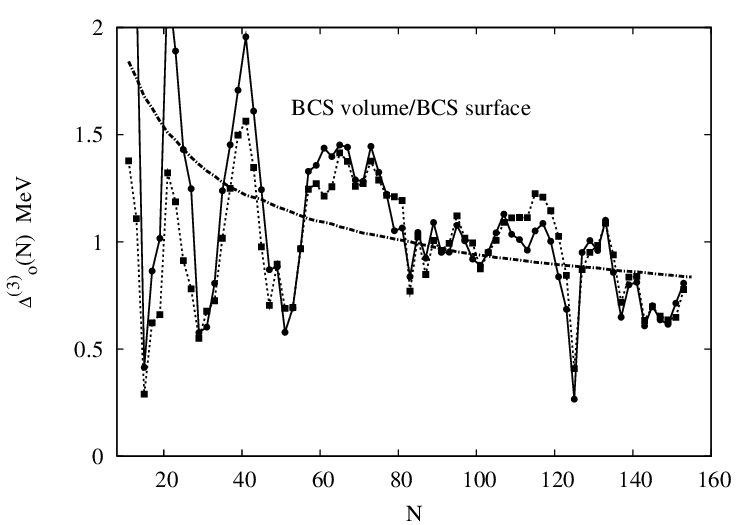}%
\caption{\label{fi:BCSvs}
Averaged BCS neutron OES with the volume (squares connected by dotted line) 
and surface (circles connected by solid line) pairing interactions of
Eq.~(\ref{deltap}).   The dot-dashed curve is the phenomenological fit to the data using the
functional form 
(\ref{Dave}).
}
\end{figure}

We have carried out the 
survey with different assumptions about the density dependence of the
pairing interaction to see sensitivity of the $A$-dependence to that
characteristic.  The results of the averaged neutron OES for volume
and surface pairing are shown in Fig. \ref{fi:BCSvs}.  The effect is very
small, except for the light nuclei.  In view of the other possible
contributions to the staggering, we do not believe that one can reliably
extract the density dependence of the effective pairing interaction 
strength from the observed $A$-dependence.  We discuss below another
mechanism that could simulate the observed trend in $A$, namely an
isospin dependence of the effective pairing interaction.

\subsection{Performance statistics}

In Table \ref{rms} we report the rms residuals for the OES in
the various treatments, fitting the strength of the pairing interaction
separately for neutrons and protons. In the case of our HFB and HFB+LN models,
we carried out separate  optimizations for neutrons and protons 
with respect to  the $x$ parameter in Eq.~(\ref{inter}).
The  effective pairing strength obtained in this way is   given by Eq.~(\ref{VHFB}).
A more proper procedure would be to
 make a two-dimensional optimization based on 
recalculated HFB mass tables assuming different strengths of proton and neutron pairing.
Our experience with HF+BCS model, however,  is that
the neutron pairing strength does not significantly affect the proton {\delo}
and vice versa, at the level of changes considered in this work. Therefore,
for the purpose of a global survey, a simplified treatment has been adopted.

From Table \ref{rms} we see that all three treatments of the pairing 
can achieve the performance of zeroth order
description as a constant (or phenomenological)
 gap, but only the Lipkin-Nogami, shown on the next-to-last line, does
significantly better.   In the HFB+LN, pairing
correlations are always present. This is particularly important for
odd-$A$ nuclei, where the standard blocking procedure often gives rise to
the unphysical pairing collapse, artificially affecting the OES and
producing an exaggerated fluctuation.  However, one should be
cautious in using the HFB+LN.
While for open-shell systems using it gives a  good agreement with those  
of the HFB 
with the full particle number projection before variation, the method
is inaccurate  for closed-shell systems \cite{[Sto07]}. Consequently, it is
safest to use HFB+LN  only away from the magic numbers.  In addition,
the numerical procedure to find the solution lacks stability when there
is a large gap at the Fermi level.  Nevertheless, we obtained converged
solutions for 440 out the 443 neutron triplets and 411 out the 418
proton triplets with our HFB+LN implementation.  The numbers in
Table~\ref{rms} are for those data sets.  If we restrict the data set
further to omit the magic numbers 28,50,82, and 126, the rms residual of
the neutron OES is hardly affected, changing from 0.23 to 0.22.

\begin{table}[htb]
\caption{\label{rms} RMS residuals of {\delo} obtained in various models.  
All energies are in MeV. The last column  shows the ratio of proton and neutron
effective pairing strengths obtained through the optimization procedure.
The mass predictions of the HFB-14 model \cite{[Gor07]} were taken from \cite{hfb14}.
 }
\begin{ruledtabular}
\begin{tabular}{lcccc}
Theory & pairing & residual&  residual & $V_0^{\rm eff}(p)/V_0^{\rm eff}(n)$ \\
 &  &neutrons   &  protons & \\
\hline
Constant  & &0.31 &  0.27 & \\
$c/A^\alpha$  & & 0.24 & 0.22 & \\
HF+BCS  &volume  & 0.31  & 0.38 & 1.05 \\
HF+BCS  &mixed  &  0.30      & 0.36 & 1.08\\
HF+BCS &surface & 0.27 & 0.35 & 1.12\\
HFB & mixed &  0.27  &  0.33 & 1.11\\
  HFB+LN & mixed & 0.23 & 0.28 & 1.11 \\
 HFB-14  & & 0.46 & 0.44 & 1.10\\
\end{tabular}
\end{ruledtabular}
\end{table}

One of the basic questions about nuclear pairing is the role of induced
interactions in the effective pairing interaction \cite{[Bri05],[Don05],[Bro06a],[Dug08],[Les08]}. 
Indirect information
about this can in principle be obtained by exhibiting the density dependence
and the isospin dependence of the effective interaction.  It is therefore
of interest to examine interactions including a density dependence to
see the sensitivity.   The rms residual for the
neutron OES with volume, mixed, and surface pairing  in HF+BCS theory are shown in Table \ref{rms}. There
is a slight favoring of the surface interaction, but we deem that the
difference in the residuals (~10 \%) is too slight to be significant.
The weak sensitivity to the density dependence confirms the results
of other studies \cite{[Dob01],[San05]}.

\subsection{Isospin dependence}
\label{isospin}
A possible isospin dependence of the effective pairing strength has 
been much discussed in the literature
\cite{[Mol92],[Vog84],[Jen86],[Sta92],[Mar07],[Yam08]}.  The nuclear
interaction may be assumed to conserve isospin at a fundamental level
but the coupling to core excitations can be different for neutron and
protons when the core has a neutron excess.  Another isospin-dependent
contribution to pairing comes from the Coulomb interaction. Indeed, 
inclusion of the Coulomb has been found to substantially suppress the
pairing interaction energy \cite{[Ang01a]} and the pairing
gaps \cite{[Les08]}.  In the last column of
the Table~\ref{rms}  we report the ratio of neutron and proton
interaction strengths we extract from our fits to {\delo}.  The
effective proton strength, needed to reproduce experimental {\delo}, is
larger than the neutron strength. If the Coulomb were included explicitly,
we would expect that the needed nuclear interaction would be even larger
for the protons.  Since the underlying strong interaction is isoscalar
to a good approximation, we believe that our inferred isospin effect 
must arise from induced three-body interactions involving the
neutron excess.  We note in passing that a number
of mass table fits by the Goriely group arrive at  pairing strengths
larger for protons than neutrons.  An example is the 
HFB-14 model \cite{[Gor07]},
shown in the last line of Table~\ref{rms}. 
However,  since  different interactions are
used for even and odd nuclei in HFB-14,  the results are not
directly comparable.

As another way to test the data for an isospin-dependent pairing
interaction, we separate the nuclei into two subsets according to neutron excess, and
compare the average residual OES.  To define the subsets, we first divide
the nuclei into 
five  $A$-bins.  For each  bin we make a cut at some
value of $I=(N-Z)/A$ to have roughly equal numbers for the two sets, which 
we designate ``low isospin" and ``high isospin''.
In that way the effect of any $A$-dependence in the OES will be
reduced.  The binning for proton and neutron values of {\delo}
 has to be done separate to
get balanced sets.  The average values of OES for the two sets
are reported in Table~\ref{ta:isospin}.
\begin{table}[htb]
\caption{\label{ta:isospin} Average {\delo}  (in MeV) calculated in HFB+LN
sorted by neutron excess.  See text for details.
}
\begin{ruledtabular}
\begin{tabular}{ccccc}
\multicolumn{2}{c}{Data set} & low isospin & high isospin & difference\\
\hline
neutrons  & exp & 1.13 & 0.94  &  -0.19\\
   & HFB+LN & 1.05 & 1.02    & -0.03 \\
\hline
protons & exp & 1.05  & 0.88  & -0.17 \\
   & HFB+LN & 0.99 & 0.93 & -0.06 \\
\end{tabular}
\end{ruledtabular}
\end{table}
The empirical OES is lower for higher neutron excesses for both
protons and neutrons.  The calculated {\delo}  for neutrons are nearly equal,
while the calculated {\delo}  for protons do show a difference but much smaller than
observed.   For both cases, the differences would require weakening the
pairing interaction for nuclei with larger neutron excesses.  An
isospin dependence would have the opposite sign for proton and neutron gaps.

As a final plausibility check on whether the different strengths could be
attributable to some isoscalar three-body interaction, we computed the
OES for nuclei on the opposite side of the $N=Z$ line (recall that our 
fitted data set is restricted to $N > Z$).  This comprises
5 neutron OES and 6 proton OES, excluding as before cases involving
$N=Z$ nuclei.  Taking the pairing strengths from our global fit, we
find that the calculated average neutron OES is low (by 0.7 MeV) while
the calculated average proton OES is too high (by 0.24 MeV).  This 
is precisely the expected direction of the error if the difference in
the effective strengths depends on the sign of $N-Z$, as would be
required by an overall isoscalar energy functional \cite{Bulgac}.  

Thus, the main evidence for an isospin dependence in the present theory
is the need for different strengths for the overall fits to neutron
and proton data sets.
This results supports the recent attempts  \cite{[Mar07],[Yam08]}  to directly parametrize 
the pairing functional in terms of isovector densities.

\section{Perspective}\label{conclusions}

  The present study demonstrates that the current state of the art in
the nuclearDFT permits calculation of OES to accuracy of the order of 0.25 MeV
rms.  This is not a trivial outcome in two respective.  First, the
binding energies involved range up to nearly four orders of magnitude
larger, so there is a high demand on computational precision in carry
out the DFT.  
Second, the pairing gap is a highly sensitive
function of the mean field properties such as level density, and
so the theory needs have an accurate treatment of the single-particle
properties.  In this work we have only considered the SLy4 functional
which has known deficiencies.  Clearly further exploration of DFT
functionals is warranted, perhaps including ones  having different
(isoscalar and isovector) effective masses 

  We found no large differences between the BCS and HFB treatments.
This is not unexpected; it is only near the drip lines that the HFB 
with its better treatment of spacial variations of the anomalous
density is needed.  On the other hand, we believe that the improvement
we found for the HFB+LN treatment is significant, showing that a 
number-conserving treatment of the pairing correlations is needed.
Many of the nuclei, particularly the odd ones,  are on the edge of 
collapse of BCS pairing, and for these a number-conserving treatment
is essential to calculate the pairing correlation energy.  Unfortunately,
the LN treatment of number violation is not reliable near closed 
shells.  We would therefore advocate in future using other
treatments of number violation, perhaps HFB with variation-after-projection
\cite{[Sto07]}, or mapping onto an exactly soluble pairing Hamiltonian
\cite{abrown}.

  The global data on OES shows a weak $A$-dependence that is
certainly not reproduced with a pairing interaction that is 
density-independent or has only a mild dependence on density.
In the calculations with the BCS theory, the sensitivity to 
density dependence was explored, and it was found that there were
only small changes in the overall performance.  It should also
be noted that the mean-field contributions to the OES
are highly dependent on $A$.  Thus, firm conclusions about the
origin of the $A$ dependence must await surveys based on theory
that avoids the filling approximation.

 A very interesting question, related to the density dependence,
is where there is an effective isospin dependence of the strength
of the pairing interaction.  The question cannot
be addressed with confidence by examining individual isotope or
isotone chains, because the other species of nucleon can affect
the effective pairing Hamiltonian, particularly the single-particle
spectrum.  However, from the global survey we find what seems to
be a robust result, that the effective pairing strength for
protons OES is about 10\% larger than for neutron OES.  The 
calculation does not take into account the Coulomb interaction 
in the pairing channel, but naively that would be expected to 
decrease the effect strength, not increase it.  The other possibility
to explain the difference is an induced isospin dependence.  

It is perhaps disappointing that the overall performance of the
theory is only slightly better than the naive one-parameter 
phenomenology attributing the staggering to a constant BCS gap.
The two-parameter phenomenological function $c/A^\alpha$ does slightly
better than the theory overall.  But that form has no justification
and the shell effects that are faithfully reproduced by the theory
are missing.  So we conclude that the rms residuals between theory
and experiment do not tell the whole story.  In any case, the promising
possibility to surpass the performance of the present phenomenology
is to continue the DFT studies with better
functionals, including mean field contributions and number-conserving
treatments of pairing.

\begin{acknowledgments}
We thank A.~Bulgac, W.~Friedman, and P.-H.~Heenen for helpful discussions. 
This work was supported in part by the U.S.~Department of Energy under Contract
Nos. DE-FC02-07ER41457 (UNEDF SciDAC Collaboration),
DE-FG02-00ER41132 (University of Washington),
DE-FG02-96ER40963 (University of Tennessee), and 
DE-AC05-00OR22725 with
UT-Battelle, LLC (Oak Ridge National Laboratory).
Computational resources were provided by the National Center for
Computational Sciences at Oak Ridge, and National Energy Research
Scientific Computing Facility.  Computations were also carried out
on the Athena cluster of the University of Washington. 
\end{acknowledgments}
 

\begin{thebibliography}{10}
\bibitem{Bulgac}  We are indebted to A.~Bulgac for calling our attention
to this point.
\end{thebibliography}

\end{document}